# Evaluation of laser diffraction-based particle size measurements using digital inline holography

Santosh Kumar.S,[1] Zilong He,[1] Christopher J. Hogan Jr.,[1] Steven A. Fredericks,[1,2] Jiarong Hong[1*]

1Department of Mechanical Engineering, University of Minnesota Twin Cities

2Winfield United, River Falls, WI

*To whom correspondence should be addressed: jhong@umn.edu

The measurements of size distribution of small particles (e.g. dusts, droplets, bubbles, etc.) are critical for a broad range of applications in environmental science, public health, industrial manufacturing, etc. Laser diffraction (LD), a widely used method for such applications, depends on model-based inversion with underlying assumptions on particle properties. Furthermore, the presence of sampling biases such as velocity differentials are often overlooked in simple ex-situ calibrations, which introduces as an additional source of error. In contrast, digital inline holography (DIH), a single camera coherent imaging technique, can both measure particle size distributions without the need for a model-based inversion and can directly provide information on the shape characteristics of the particles. In this study, we evaluate the performance of an LD system in characterizing polydisperse droplets produced in a flat fan spray using in-situ DIH based imaging as a reference. The systematic differences in the two techniques are examined. A droplet-trajectory-based correction for the LD-inferred size distributions is proposed to compensate for the observed differences. We validate the correction using NIST standard polydisperse particles undergoing differential settling, and then apply the correction to polydisperse spray droplet measurements. The correction improves agreement between LD and DIH size distributions for droplets over two orders of magnitude, but with LD still underestimating the fraction of droplets at sizes above ~1 mm. This underestimation is possibly linked to the complex oscillatory and rotational motion of droplets which cannot be faithfully captured by measurement or modelled by the correction algorithm without additional information.

## 1. Introduction

The measurement of particle size distributions (e.g., bubbles, droplet, sediments, etc.) is critical in characterizing and predicting the behavior of many natural and industrial processes. For example, assessing the impact of atmospheric aerosol particles from sources including sea spray, volcanic activity, and dust, on climate, requires careful analysis of particle size distributions [1]. In industrial practices including but not limited to the spray drying of food products [2], spray-based application of crop protectants [3], direct fuel injection for combustion [4], drug manufacturing [5,6], and multiphase chemical reactors [7], the size distribution of particles is often a critical process control parameter that requires careful and continuous monitoring.

Laser diffraction (LD) has been established as a standard technique for particle size distribution measurement, namely because it utilizes a relatively large sampling volume, has a high sampling rate, and with implementation of automation, has a straightforward operation procedure [8,9]. LD employs a collimated laser beam to illuminate a group of particles. Each particle within the group produces a characteristic angular intensity pattern through forward scattering, the ensemble of which is captured on a radial sensor. Once captured, the intensity can be numerically inverted to obtain the size distribution for the group of particles. Inference of the particle size distribution from the radial scattering distribution function requires use of an ill-posed inversion, in which it is commonplace to assumes particles are spherical, with light scattering following either Fraunhofer or Mie theory, the latter of which requires the refractive index of the sample to be known *a priori* [10]. LD has been successfully applied in measuring size distribution of various types of particles in a wide range of industries, including food processing [11], agriculture [12], paints & coatings [13], manufacturing [14], oil & gas [15] and pharmaceuticals [16].

Despite its widespread use for size distribution measurement, LD suffers from several limitations. First, the assumption of spherical shape and refractive index (for Mie theory) can act as a significant source of error for the measurements of non-spherical or irregular particles. For instance, Agimelen et al. (2017) found that the presence of needle shaped particles introduces multimodal populations in inverted results with modes arising which are smaller than the actual size of particles [17], while Andrews et al. (2010) showed that measurements of a mixture of organic and inorganic particles with different refractive indices yielded poorer agreement with true distributions than measurement of either sample independently [10]. Second, the use of a fixed-geometry radial sensor results in poor resolution, limiting the ability of the technique to differentiate narrow size distributions [18]. Such a limit in resolution is more apparent for particles of larger sizes as their forward scattering energy is restricted to regions of smaller angles [19]. Beyond these potential issues, for particles in motion, preferential weighting of slower velocity samples (over-counting) relative to faster ones can lead to a size-dependent sampling bias if particle velocity is a function of size [20–22]. Such biases can be more apparent in spray measurements, where the nozzle is often vertically traversed for an ensemble measurement across the spray width [12] and where droplet inertia can strongly affect its velocity. Furthermore, the presence of spatial variations due to differences in spray breakup mechanisms along and away from the centerline [23,24], as well as droplet oscillations, can act as additional sources of error when interpreting LD results.

Several studies present correction strategies for mitigating the above limitations of LD based measurements. Heffels et al. (1995) introduced a modified inversion algorithm for non-spherical particles, but its implementation requires knowledge on the particle aspect ratio [25]. To overcome the spatial sampling bias, Fritz et al. (2014) proposed to increase background wind speed as a method to diminish the impact of differential velocities, which is largely limited to controlled laboratory environments [21]. There is, however, limited applicability to these correction strategies. Furthermore, the calibration of LD systems traditionally utilizes either a direct measurement with spherical particle or reticle standards [26,27], or comparison to secondary measurements such as sieving [11,28], coulter counters [29,30], sedimentation [31], or optical/electron microscopy [32], all of which are performed under ex-situ conditions. To date, an in-situ calibration approach has not been established for LD, though this would be extremely beneficial as the application of spherical particle/reticle standards fails to capture the complexity seen in measurements e.g., droplet oscillations in sprays causing asphericity [33], and the use of secondary measurements may introduce modification of the particles under examination (through aggregation or dispersion).

In contrast to the noted techniques, imaging-based approaches enable high resolution model-free direct sizing of arbitrary shaped particles located in a large sampling volume with single object sensitivity and can further eliminate the spatial sampling bias with an appropriate choice of frame rate. In particular, digital inline holography (DIH) has recently emerged as a versatile tool for characterizing samples in-situ at high spatial resolution and an extended depth of field (typically more than 3 orders higher than conventional imaging) using a single camera [34]. DIH captures the interference pattern (i.e., hologram) between the laser light scattered by the particles and the unscattered portion of the beam, which encode the 3D position and shapes of the detected object. Once recorded, the hologram can be numerically reconstructed based on different diffraction formulations (e.g. Rayleigh-Sommerfeld or Kirchhoff-Fresnel formulations), providing the complete 3D optical field containing the particles. Evolution of the size, shape and position of the particles can be subsequently extracted and tracked over time using standard image processing algorithms [35]. Furthermore, as an imaging-based measurement, DIH only requires calibration of the pixel resolution using precision target with known dimension, without any other secondary measurements or standardized particle samples. The high level of accuracy and sensitivity provided by DIH has enabled multiple applications, including measurement of snowflakes [36] and droplets [37] in the atmosphere, sediments [38] and oil droplets [39] in oceans, coal particles in flames [40] and bubbles in the wake of a ventilated supercavity [41]. More recently, Kumar et al. (2019) demonstrated the versatility of DIH to fully resolve monodisperse droplets generated from a vibrating orifice aerosol generator as well as from polydisperse flat fan sprays [33]. Along with the standard size distribution, the study also quantified the particle shape characteristics using a volumetric size-eccentricity joint probability density function (PDF), a quantity that is typically challenging to obtain through LD measurements.

Given the widespread reliance of LD on ex-situ calibration, there still exists a clear need to fully assess its accuracy with an in-situ calibration approach, in order to identify potential sources of error and the corresponding mitigation strategies. In this study, we perform a systematic evaluation of LD in characterizing the droplet size distribution of a polydisperse flat fan agricultural spray using an in-situ high-fidelity DIH based imaging approach as a reference, with both measurements performed at identical experimental conditions. A detailed description of the experimental setups and measurement conditions are presented in Section 2. In Section 3, we present a comparison of the two measurements, a proposed trajectory-based correction, validated using polydisperse NIST standard beads, and a comparison of the corrected LD distribution with the DIH data. Finally, we conclude with a summary and discussion in Section 4.

## 2. Experimental Methods

In this section we provide a detailed description of the experimental setup used to generate the spray test cases, the LD and DIH systems deployed to characterize the size distributions of spray droplets. All of our experiments are conducted in a low speed recirculating wind tunnel (Hambleton Instruments, Hudson WI) with a 3.20 m long, 0.91 m wide, 1.83 m tall test section, capable of achieving wind speeds up to 8 m/s. The test section (Fig. 1a) has clear glass walls providing optical access from the sides. A water spray (tap water at 19 ºC) from a TP6515 flat fan nozzle (65° fan angle; major axis diameter $D_N \sim 4.1$ mm, TeeJet Technologies) at a pressure of 152 kPa (measured across the nozzle) is introduced in the tunnel, oriented along the centerline and parallel to the flow direction, under an air speed of 4 m/s. The measured liquid feed rate (using a Coriolis Mass Flowmeter, RCT1000, Badger Meter Inc.) under these conditions is 4.08 liters per minute, in good agreement with ASABE S572.2 standards [42]. The nozzle is attached to a vertical translation system (similar to [12]) that permits measurements across the entire span of the spray fan. The tunnel is equipped with a mist eliminator downstream of the test section to prevent droplets from recirculating in the tunnel. The temperature and humidity inside the test section are continuously monitored during measurements at 30 °C and ~80%, respectively, to ensure identical experimental conditions across all measurements. The spray droplet size measurements are conducted in four individual sampling locations within the spray fan as shown in Figure 1b with detailed locations and dimensions summarized in Table 1. The sample locations are selected considering the droplet concentration range in which both LD and DIH can operate and the need to evaluate the spatial variation of droplet size distribution and its impact on the LD measurements.

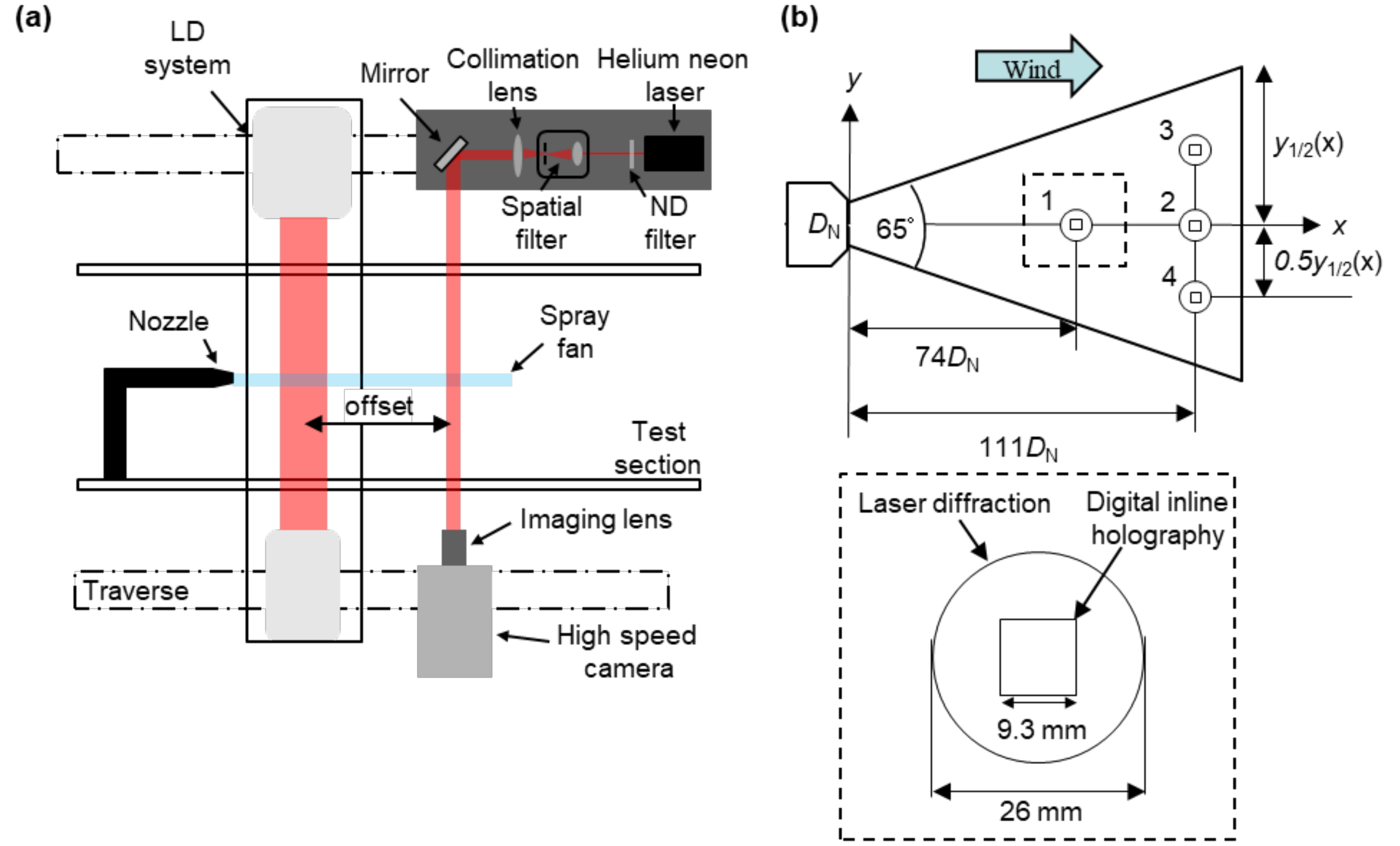

**Figure 1.** (a) A schematic diagram of the test section indicating the laser diffraction (LD) system on the left and the digital inline holography (DIH) system on the right . (b) A Schematic of the spray fan from a TP6515 flat fan nozzle with the four sampling locations marked, with an inset illustrating the field of view for laser diffraction (LD) and digital inline holography (DIH). The arrow at the top indicates the direction of flow in the wind tunnel from left to right and $D_N$ and $y_{1/2}(x)$ are the diameter of the nozzle and the jet half-width respectively.

The DIH system (Fig. 1a) consists of a 12 mW helium-neon laser (REO Inc.), a neutral density (ND) filter to control the laser intensity, a spatial filter (Newport Inc.) to increase laser spatial coherence, a collimation lens with 75 mm focal length (Thorlabs Inc.) and a high speed camera (Phantom v710) with an imaging lens (Nikon 105 mm f/2.8), all mounted on either side of the test section. The optical components together produce a 50 mm collimated gaussian beam which is captured on a 512x512 pixel image at a resolution of 18.2 µm/pixels by the camera. The calibration involves capturing an in-focus image of a precision microruler with 10 µm spacing over a 1 mm range (Thorlabs Inc.; R1L3S2P) as described in [33]. A frame rate of 500 frames/s ensures that no droplets are sampled more than once, avoiding the spatial sampling bias of slower droplets on the distribution. Images are recorded for a duration of 1.6 minutes, yielding ~49000 holograms in total at each individual sampling location. The collected holograms are processed using an automated reconstruction and sizing routine to extract the size distributions as well as the size-eccentricity joint PDF, the complete details for which can be found in [33]. The study includes the validation of the imaging accuracy, by calibration with a precision microruler (Thorlabs Inc.; 10 µm line spacing), and the measurement precision, defined by the peak to width ratio of 14.2, measured with monodisperse droplets produced by a vibrating orifice aerosol generator.

**Table 1.** Position of the sampling locations on the TP6515 nozzle generated flat fan spray along with their corresponding symbols used in the size distribution plots, with $D_N$ and $y_{1/2}(x)$ representing the nozzle diameter and the half width of the jet respectively.

| | Streamwise Position | Spanwise Position | Symbol |
|---|---|---|---|
| Position 1 | $74D_N$ | 0 | ○ |
| Position 2 | $111D_N$ | 0 | □ |
| Position 3 | $111D_N$ | $0.5y_{1/2}(x)$ | △ |
| Position 4 | $111D_N$ | $-0.5y_{1/2}(x)$ | ◇ |

The LD system (Fig. 1a) employed in our experiment is a Helos/KR-VARIO laser diffraction system (Sympatec) with a 2000 mm focal length lens, placed across the glass windows of the test section, for measurements in the 18-3750 µm size range. The software suite captures and analyzes the data for all our experimental cases from a sampling window of ~26 mm and includes an independent referencing step, without the spray, before each measurement. The total sampling duration for each measurement is 20 s with a signal integration time of 5.8 ms. We further perform 10 replicates for each sample to ensure size distributions are stable and converged. The DIH and LD systems are mounted together on a traverse, with a constant lateral offset between them, which enables us to move the sampling locations for both in the downstream direction. In combination with the vertical nozzle translation, this lateral motion enabled measurements at any arbitrary location within the entire spray fan.

## 3. Results & Discussion

3.1 Droplet size distribution comparison: laser diffraction *vs* digital inline holography

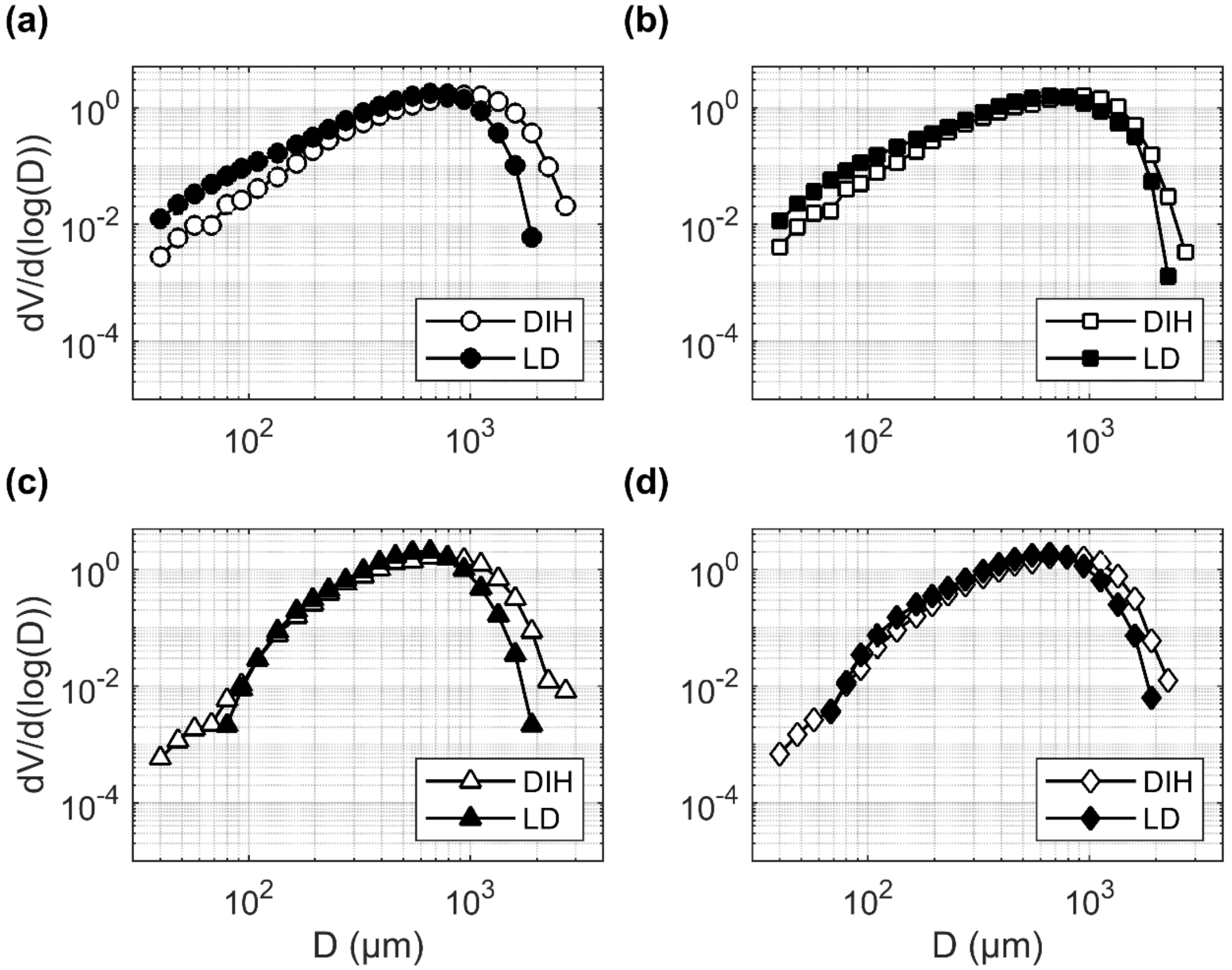

**Figure 2.** Volume-based size distributions for spray droplets generated by a TP6515 flat fan nozzle measured by digital inline holography (DIH, open symbols) and laser diffraction (LD, closed symbols) on a log-log plot. Comparisons between the two techniques are performed at **(a)** $74D_N$ (position 1) and **(b)** $111D_N$ (position 2) downstream of the nozzle along the centerline and at $111D_N$ downstream and **(c)** $0.5y_{1/2}(x)$ above (position 3), **(d)** $0.5y_{1/2}(x)$ below (position 4) the centerline, where $D_N$ and $y_{1/2}(x)$ are the nozzle diameter and half width of the jet at the measurement location, respectively.

Figures 2 presents a comparison of the droplet size distributions on a volume basis for both laser diffraction (LD) and digital inline holography (DIH) at the four sampling locations, with identical logarithmically spaced bins. The distributions have been normalized resulting in the integral area under the PDF of unity. The distribution at position 1 (Figure 2a) is monomodal with similar polydisperse shapes resulting from both LD and DIH measurements. However, the LD distribution shows a clear mismatch with the DIH result for all sizes. Despite a larger sampling window for LD compared to DIH, it fails to capture any droplets in the largest diameter sizes due to the limited spatial resolution of a radial LD detector as reported by [18]. Furthermore, a larger FOV often associated with a loss of resolution actually measures an increase at smaller sizes compared to DIH possibly due to oversampling those droplets which tend to move slowly. Although the measurement window sizes are different, their size relative to the spray fan (~1-6 % of the spray fan width) is significantly smaller and thus would not capture any spatial variation in droplet size within them. In addition, the geometric mean diameter obtained through a lognormal fit of the LD distribution is ~200 μm smaller than that from DIH (see Table 2 and supplementary information), further illustrating the difference between the two more clearly. For measurements downstream at position 2, the shapes of the individual distributions, presented in Figure 2b, remain the same as in position 1, but with a decrease of ~90 μm and ~13 μm in the corresponding geometric mean diameters for DIH and LD, respectively. The observed decrease may be caused by droplet breakup and evaporation as droplets migrate downstream, a trend which was observed and reported on for similar measurement conditions by [33]. The differences between LD and DIH results are suppressed across all sizes at position 2, with the variation of geometric mean diameter between the two reducing to ~130 μm. Likewise, the trend in underestimation of larger sizes, the failure to capture any droplets in the largest diameter bin and the overestimation of smaller sizes by LD continues to hold at this position.

As the sampling location shifts spanwise above the centerline to position 3, the distributions for LD and DIH continue to be monomodal and highly polydisperse (figure 2c), but with further reduction in geometric mean diameters by ~40 μm and ~75 μm for the LD and DIH results relative to position 2, respectively. The sharp decline in the relative concentration of smaller diameters and the decrease in the geometric mean diameter are due to the variation in the break up mechanism away from the centerline [24]. Apart from the loss of the largest diameter droplets in LD, caused by dynamic range limits, the lower concentration of small droplets below ~80 μm leads to no perceptible signal in the detector due to the weak scattering strength not surpassing the minimum optical concentration required [43]. Such a limitation however does not exist for image-based DIH measurement, which is characterized by high measurement sensitivity and particle sizing resolution [33]. The match between the LD and DIH size distributions is improved between ~200 μm and ~800 μm, but underestimation in LD above the modal peak results in a geometric mean diameter difference of ~90 μm, with LD still less than DIH.

Finally, at position 4, LD and DIH distributions (figure 2d) show an increase of ~10 μm and ~20 μm respectively, in the geometric mean diameter relative to position 3. Although noteworthy, the increase in the geometric mean diameter is still leads to smaller values than are measured at position 2, ensuring a consistent trend at both off-center positions compared to the centerline. We believe this increase is caused by effect of gravity introducing additional variations in breakup at the bottom of the spray relative to the top, and is reliably captured by both LD and DIH measurements. This effect further increases the deviation between LD and DIH for diameters between ~90 μm to ~250 μm with the difference in geometric mean diameter between LD and DIH increasing to ~110 μm from, as compared to ~90 μm at position 3. The failure of LD in capturing droplets in the largest size bins, droplets smaller than ~80 μm, and the underestimation of the PDF above the modal peak also persist at this measurement location.

**Table 2.** Comparison of geometric mean and geometric standard deviation for lognormal fits of the laser diffraction (LD) and digital inline holography (DIH) based size distributions.

| | LD | | DIH | |
|---|---|---|---|---|
| | Geometric mean (μm) | Geometric std. | Geometric mean (μm) | Geometric std. |
| Position 1 | 617.5 | 1.66 | 821.9 | 1.76 |
| Position 2 | 603.7 | 1.78 | 732.3 | 1.80 |
| Position 3 | 561.5 | 1.56 | 658.2 | 1.75 |
| Position 4 | 570.9 | 1.64 | 680.1 | 1.73 |

In total, measurements at all four locations consistently show that the droplet size distribution from LD is peaked at smaller sizes than that directly determined from DIH, irrespective of variations in the droplet size distribution at the various positions [23]. The difference in LD may be attributed to the presence of a size-dependent sampling bias as well as ambiguity introduced by non-spherical droplets in the sample. As the droplets move through the sampling location, size-dependent differences in the speeds, caused by different drag behaviors, would lead to overcounting of the slower moving droplets [21,22], while the presence of non-spherical droplets can introduce an error in model inversion of LD which assume a spherical shape. In order to better characterize the level of asphericity in the sample, we investigate the size-eccentricity joint PDF obtained from DIH measurements in the following section.

3.2 Size-eccentricity joint PDF from digital inline holography

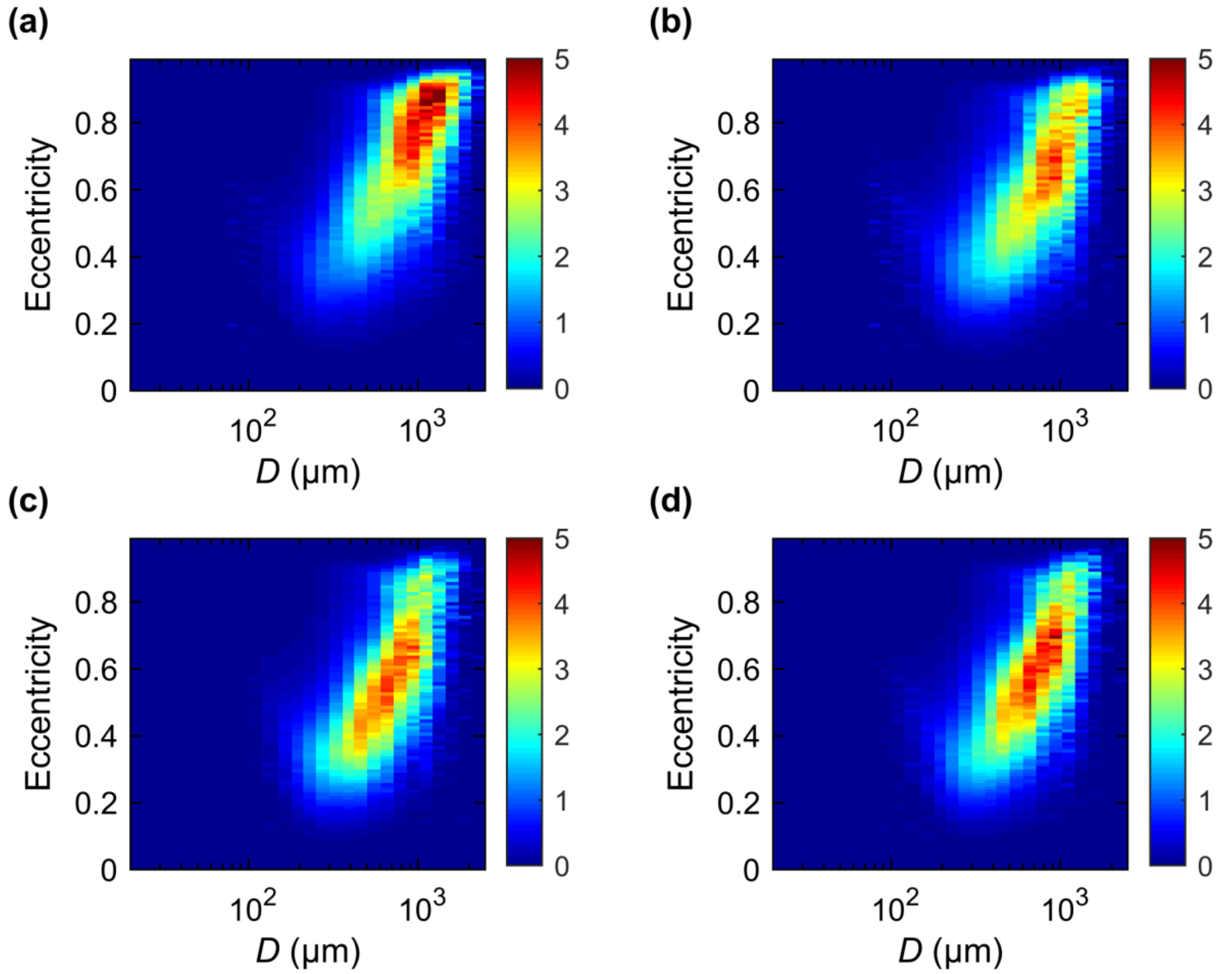

**Figure 3.** The volumetric size-eccentricity joint PDF for TP6515 flat fan spray generated droplets measured at the four sampling locations. The two position along the centerline at **(a)** $74D_N$ (position 1) and **(b)** $111D_N$ (position 2) downstream of nozzle. The two off-center spanwise positions at $111D_N$ downstream and **(c)** $0.5y_{1/2}(x)$ above (position 3) and **(d)** $0.5y_{1/2}(x)$ below (position 4) the centerline where $D_N$ and $y_{1/2}(x)$ are the nozzle diameter and half width of the jet at the measurement location, respectively.

The volumetric size-eccentricity joint PDF of the flat fan spray at the four positions obtained through DIH measurement is presented in Figure 3. The eccentricity is defined by $\sqrt{(1-(a^2/b^2))}$ with $a$ and $b$ as the semi-minor and semi-major axes of the ellipse, respectively. They are fit to the cross section of the droplets captured by DIH. In addition, the calculated eccentricities are weighted by volume so that an integration over all eccentricities will recover the volumetric size distribution presented earlier (see [33] for more information). The contours of the PDF reveal a strong semilogarithmic scaling between eccentricity and diameter at all positions. Such a scaling can be rationalized by the presence of droplet oscillations and rotations, which are illustrated by snapshots of high speed shadowgraphy shown in Figure 4, the videos and the experimental details for which are included in the supplementary information. Note that the PDF's are a statistical measurement of all droplets that cross the sampling window rather than an instantaneous eccentricity. Thus, an apparent drop in the eccentricity joint PDF above ~1 mm, seen in Figure 3b, c and d, do not indicate that larger droplets are more spherical, but rather are indicative of the lack of such large droplets at these locations. Specifically, the pinch off of droplets from the liquid lamella initiates oscillations along the direction of motion (Figure 4a) which are driven by surface tension effects, while rotations (Figure 4b) are caused by the moments associated with wind-induced drag on the droplet. Once initiated, the oscillations decay due to viscous dissipation, the time scale of which scale inversely with diameter causing smaller droplets to relax faster than larger ones. On the other hand, rotational motion leads to instabilities of droplet shape resulting in further breakup of the droplet [44]. The PDF at position 1 (Figure 3a) indicates a strong peak around ~2 mm and 0.9 eccentricity which shifts to a smaller diameter (~0.9 mm) and eccentricity (~0.7) at position 2 (Figure 3b). Such a peak reduction is caused by the relaxation of droplet oscillations as well as breakup of droplets undergoing rotation as they travel downstream, the latter of which is evidenced by the size distributions presented in Section 3.1, and previously reported under similar experimental conditions in [33].

At both the off-center positions (Figure 3c and 3d), the PDF envelope retains the same shape as previous locations, but with a significantly broader peak spread spanning a diameter between 0.4 mm to 1 mm and eccentricities between 0.5 to 0.7. Apart from a similar range of rotational and oscillatory motion exhibited by droplets along the centerline, the angular difference between the direction of wind and droplet motion in the off-center position leads to oscillations along multiple directions, oriented with the major and minor axes of the droplet, as it moves (Figure 4c). The presence of such three dimensional oscillations increases the effective oscillation frequency of the droplet [45], reducing the total relaxation time required for droplets to reach equilibrium. As a consequence, both off-center positions have larger fraction of droplets exhibiting smaller asphericities relative to the centerline, broadening the measured PDF along eccentricity. Furthermore, the complex three-dimensional oscillations also result in breakup of larger droplets, as described in Section 3.1. This causes the observed PDF spread along diameter. Finally, because of gravitational settling, there is a slight shift in the peak towards larger diameters and eccentricities at the bottom of the spray fan (position 4) relative to the top (position 3).

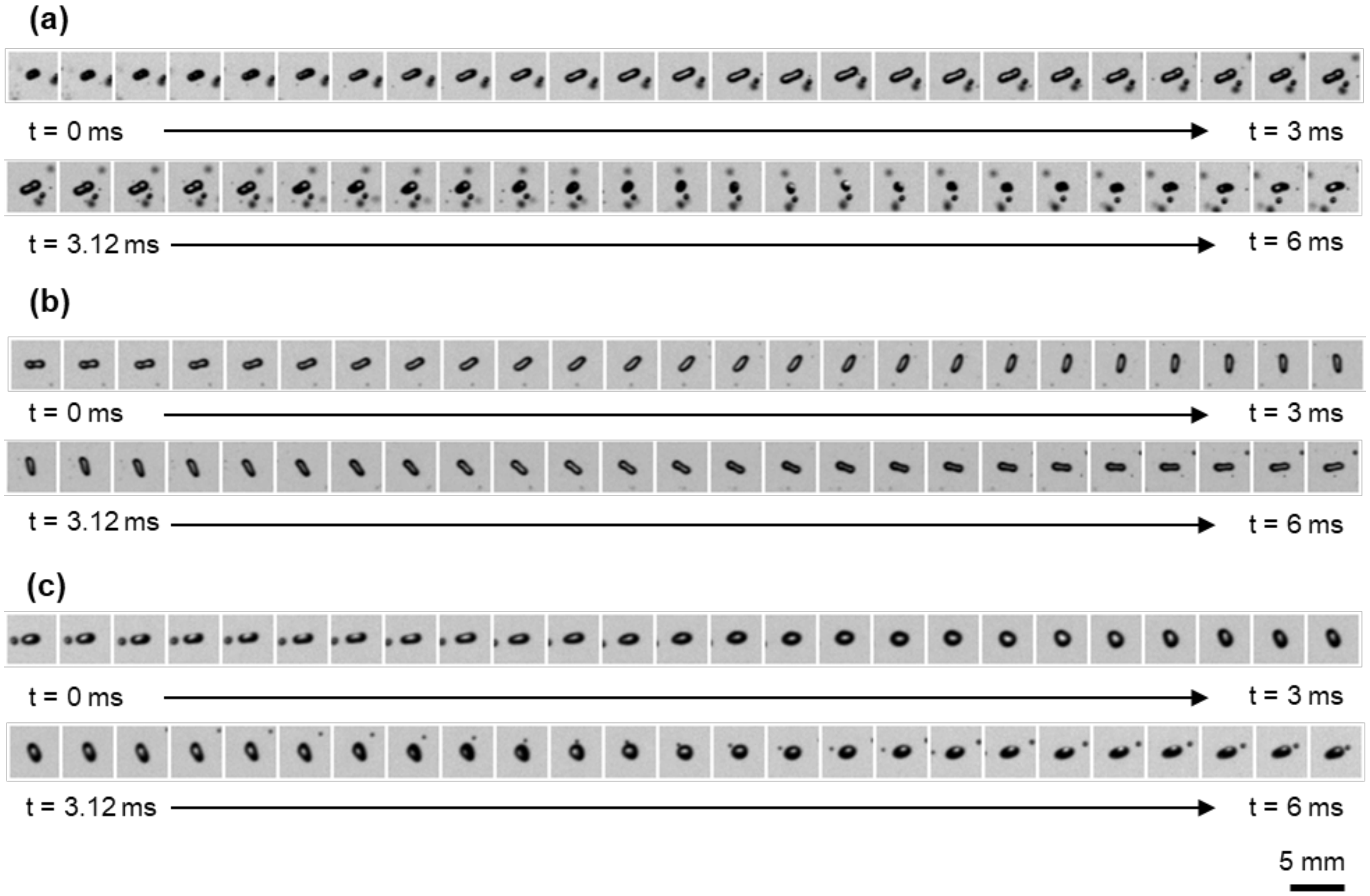

**Figure 4.** Snapshot montage from high speed shadowgraphy of complex droplet dynamics within the spray exhibiting **(a)** oscillations along direction of motion **(b)** counterclockwise rotation with axis perpendicular to image plane and **(c)** three dimensional oscillations along the major and minor axes of the droplet. The time interval between snapshots are 0.12 ms. The images are presented in a Lagrangian frame of reference centered on each droplet

3.3 Trajectory-based correction of LD results

In order to compensate for the observed differences between the LD and DIH measurements, we propose a correction based on the transit time of droplets crossing the measurement sampling volume. The liquid velocity at the nozzle exit is significantly higher than the wind velocity, and droplets initiated by liquid jet breakup tend to move at this velocity. However, the difference between the wind speed and droplet velocity results in a net drag force acting to slow down droplet velocities. The smaller droplets, of lower inertia, tend to reach terminal velocity quicker than larger droplets, resulting in a differential velocity based on size and position. As the droplets cross the laser beam of the LD system, the slow-moving smaller diameter droplets are counted more often than the faster moving larger droplets, leading to a sampling bias as described in Section 1. In contrast, DIH data, which is sampled at a suitable rate to ensure each droplet is imaged only once, avoids this problem. A simple approach to correct this bias involves the application of the Verlet Algorithm [46], to estimate the transit time of a droplet as it crosses the LD sampling region, which can then be used to adjust the droplet count to limit the effect of the bias. The algorithm, given by the equation motion for a droplet in air (equation 1) can be used to estimate the time spent by it when crossing the laser beam as a function of diameter:

$$\vec{F} = m\overrightarrow{a_n} = -\frac{\frac{1}{2}C_d\rho_{air}\pi d_p^2}{4}\left|(\overrightarrow{v_n} - \overrightarrow{u_{air}})\right|(\overrightarrow{v_n} - \overrightarrow{u_{air}}) + m\vec{g} \quad (1)$$

$\overrightarrow{v_n}$, $\overrightarrow{a_n}$ are the velocity and acceleration of the droplet at time $n$, $m$ the mass, $C_d$ the drag coefficient, $\rho_{air}$ the density of air, $d_p$ the diameter of the droplet, $\overrightarrow{u_{air}}$ the wind velocity and $\vec{g}$ the acceleration due to gravity. In order to incorporate the effect of droplet rotation and oscillations, we model droplets as ellipses with the aspect ratio obtained from the ridge lines of the size-shape joint PDF (Figure 3). Next we combine the drag model proposed by [47] for non-spherical inertial solid particles moving with random orientation with a correction for liquid deformation proposed by [45]. The combination of both models in our understanding is unique and has not been reported before. The complete equations for the drag model can be found in the supplementary information. The wind velocity is approximated as a uniform flow in the x-direction at 4 m/s; we note that this approximation does not enable considering of vertical, shear-induced, aerodynamic focusing.

The model is initialized by assuming the initial droplet velocity ($\overrightarrow{v_{init}}$) to be uniform at the exit of the nozzle and given by the ratio of the flow rate to nozzle area, ignoring any interaction between droplets and spans the entire angular spread of the jet i.e., 65°. Next the droplets are stepped in time with no external forces, up to a minimum distance of ~$40D_N$ to simulate the formation and breakup of the liquid sheet. Once outside the sheet, the droplets are stepped in time with a step size of $10^{-4}$ s, sufficient to fully resolve the motion of all diameters. We count the time each droplet resides within the measurement window of the LD system (a circle of 32 mm diameter) placed at each corresponding sampling locations. The correction factor $C$ (equation 2) is defined as the inverse ratio of total transit time divided by the time for the largest (quickest) droplet to cross the laser as a function of droplet size ($d_p$).

$$C(d_p) = \frac{t(d_p)_{min}}{t(d_p)} \tag{2}$$

Once calculated, we multiply this monotonically increasing non-dimensional function, with a value of one at the largest size, to the measured LD size distribution and renormalize it to eliminate the sampling bias present in the data. Normalization of the PDF involves dividing the number of samples in each bin by the logarithmic bin width which results in the integral area under the curve to be unity.

3.4 Validation of trajectory-based correction

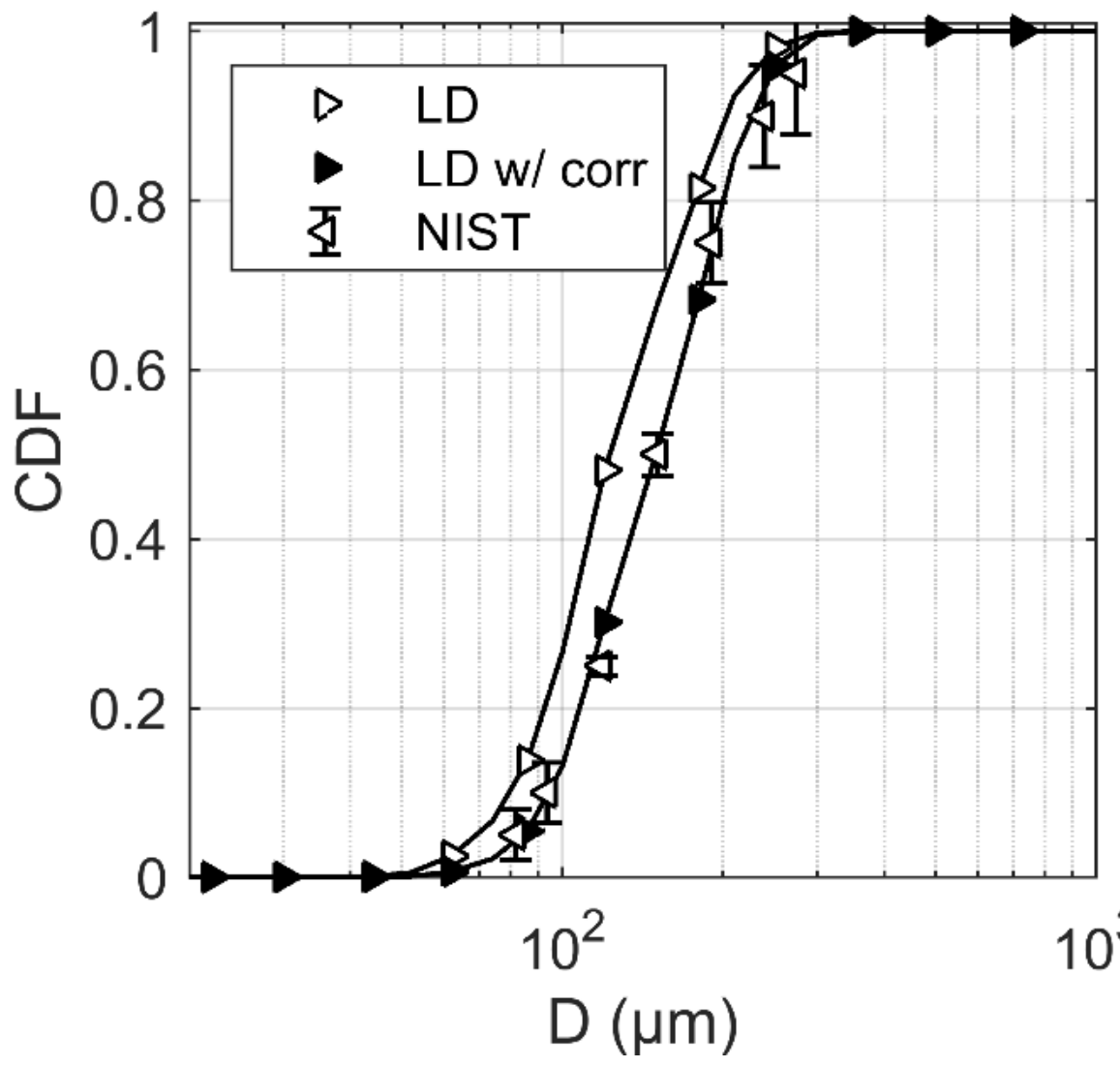

**Figure 5.** Comparison of cumulative distribution functions (CDF) of NIST polydisperse particles measured with laser diffraction (LD) compared to the corrected laser diffraction (LD w/ corr) and the NIST standard results with error bars indicating measurement uncertainty.

We validate our correction algorithm by measurements of NIST standard spherical polydisperse beads (soda lime glass) in the 50-350 µm range, from Whitehouse Scientific, using the laser diffraction (LD) system [48]. The particles are collectively dropped from above the LD sampling volume through the laser and collected on the bottom resulting in a differential settling velocity based on size. The measured volumetric cumulative distribution function (CDF) shown in Figure 5 (resulting from integration over the entire drop-time) clearly illustrates this sampling bias as an underestimation relative to the NIST standard, with an error of ~30 µm at 50% volume. On application of the trajectory-based correction, we eliminate the effect of settling velocity on the distribution and obtain a closer agreement to the NIST standard distribution over the entire size range of the measurement. With the approach validated, we will next apply the algorithm to correct LD measurements of the TP6515 flat fan spray generated droplets.

### 3.5 Droplet size distribution comparison: corrected laser diffraction *vs* digital inline holography

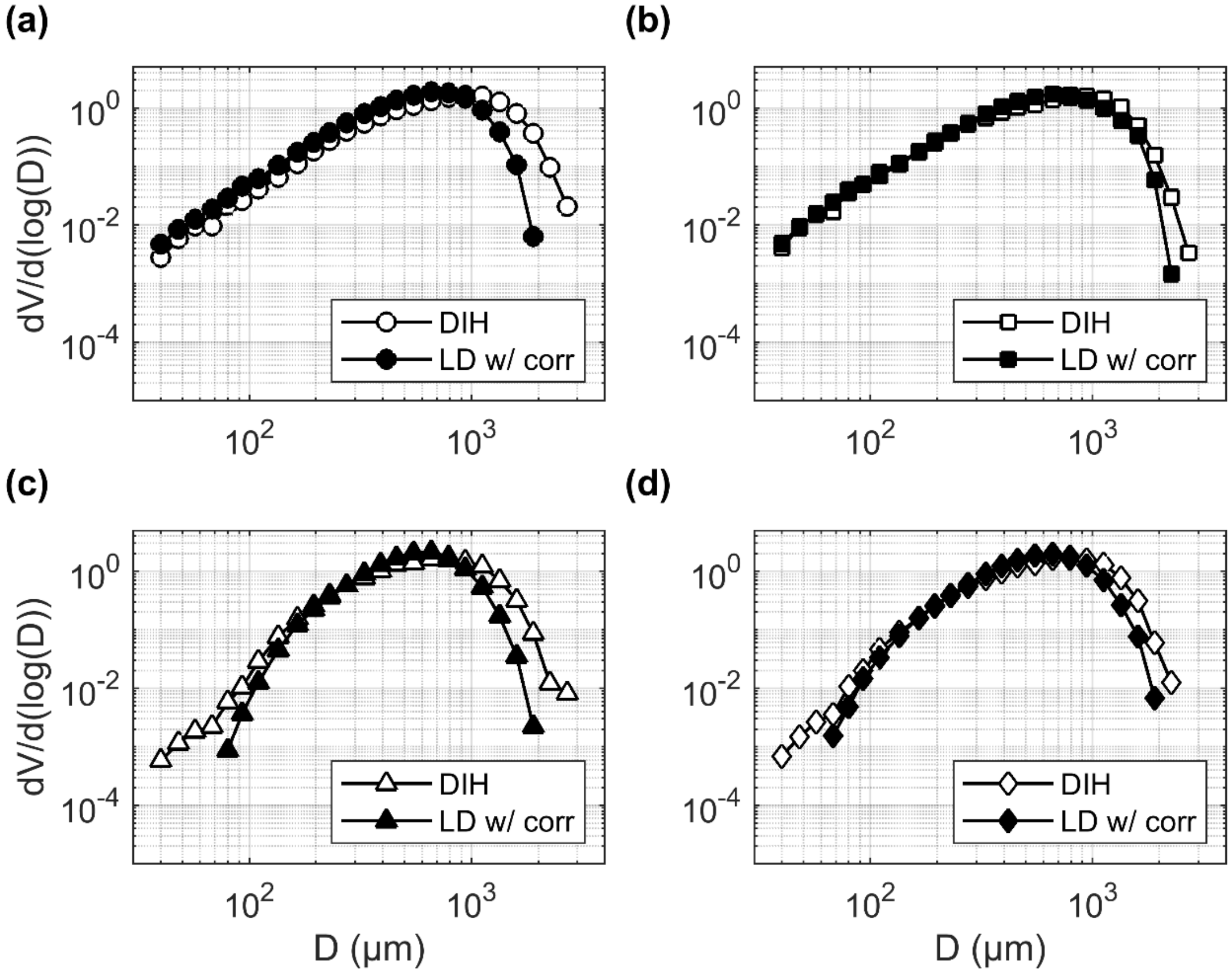

**Figure 6.** Volume-based size distributions functions for spray droplets generated by a TP6515 flat fan nozzle measured by digital inline holography (DIH, open symbols) and corrected laser diffraction (LD w/ corr, closed symbols) on a log-log plot. Comparisons between the two techniques are performed at **(a)** $74D_N$ (position 1) and **(b)** $111D_N$ (position 2) downstream of the nozzle along the centerline and at $111D_N$ downstream, **(c)** $0.5y_{1/2}(x)$ above (position 3), and **(d)**

$0.5y_{1/2}(x)$ below (position 4) the centerline, where $D_N$ and $y_{1/2}(x)$ are the nozzle diameter and half width of the jet at the measurement location, respectively.

We apply the validated trajectory-based correction to the laser diffraction (LD) data and compare the volumetric size distributions obtained to corresponding values from digital inline holography (DIH) which are presented in Figures 6. The corrections result in the reduction of the LD distribution values across all diameters, similar to the NIST calibration; hence LD-based size distributions are renormalized following application of the correction factor. After correction application, at position 1 (Figure 7a), the discrepancy between the LD and DIH decreases for all sizes below ~800 μm, but an underestimation at larger sizes still persists, and the correction only results in a marginal drop in the geometric mean diameter difference to ~170 μm. At position 2 (Figures 6b), the agreement between the two measurements show significant improvement compared to position 1, with a near perfect overlap across all sizes below the peak of the PDF. Interestingly, the mismatch at the larger diameters show no significant change, owing to the fact that the correction factor is close to unity at these sizes. The geometric mean diameter difference between the two techniques also drops to ~100 μm from ~130 μm measured before the correction. We suggest the improved performance of the correction can be linked with a decrease in asphericity of droplets at position 2 represented by the peak shift in the size-shape joint PDF (Figure 4b).

Moving to the off-center location above the centerline at position 3 (Figure 6c), correction application only leads to a marginal change. In addition, apart from the loss in smaller diameter particles which cannot be recovered using a multiplicative correction, the under-counting at larger sizes also remains, leading to a geometric mean diameter difference of ~80 μm being retained. Finally, at position 4 the two distributions continue the trend seen at other positions. Along with the increased range of agreement between LD and DIH after correction, the mismatch observed between ~90 μm and ~250 μm is also suppressed. While correction application certainly improves agreement between LD and DIH-inferred size distributions, across all positions, we observe the correction fail to eliminate the underestimation in the droplet counts at diameters above the mode diameter. We suggest this is attributable to the complex morphology of droplets present at sizes approaching and exceeding 1 mm, as these droplets undergo three dimensional oscillations and rotations, as illustrated in Figure 4, which the proposed drag model does not account for precisely, as it only relies on a planar elliptical description of the droplets. Unfortunately, even with the 3D imaging capabilities of DIH, we are still unable to measure deformation perpendicular to the image plane, which can be obtained by employing two orthogonal DIH systems imaging the same field of view. We hence suggest judicious interpretation of LD measurements in the millimeter size range for deformable objects, such as liquid droplets.

**Table 3.** Comparison of geometric mean and standard deviation for log-normal fits of the corrected LD and DIH size distributions

| | LD (Corrected) | | DIH | |
|---|---|---|---|---|
| | Geometric mean (μm) | Geometric std. | Geometric mean (μm) | Geometric std. |
| Position 1 | 630.3 | 1.62 | 821.9 | 1.76 |
| Position 2 | 638.0 | 1.70 | 732.3 | 1.80 |
| Position 3 | 578.6 | 1.53 | 658.2 | 1.75 |
| Position 4 | 592.0 | 1.59 | 680.1 | 1.73 |

3.6 Effectiveness of physics-based correction

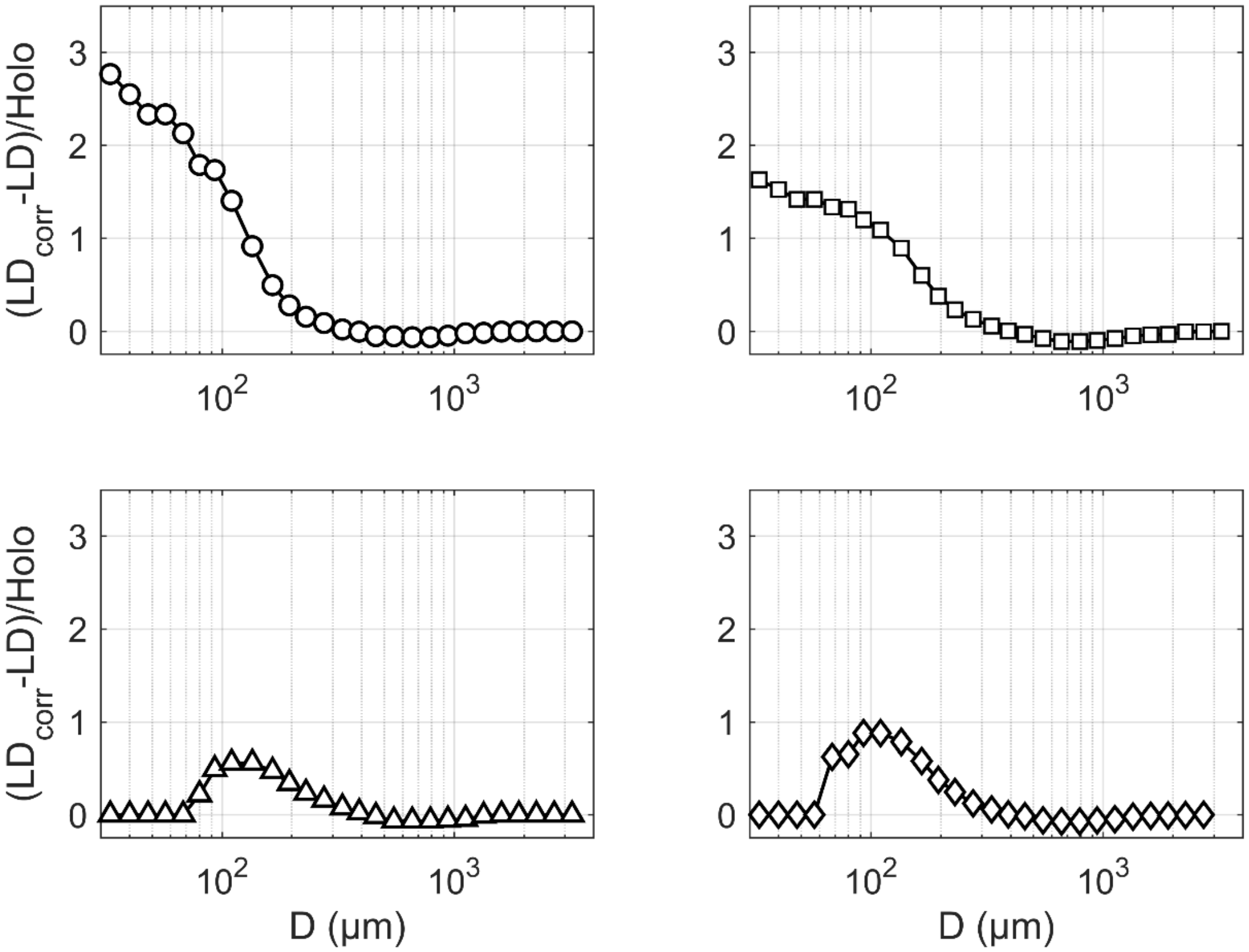

**Figure 7** Relative error between corrected and original laser diffraction (LD) measurements with respect to digital inline holography (DIH) based size distribution for TP6515 flat fan nozzle spray droplets at distances of $74D_N$ and $111D_N$ downstream of the nozzle corresponding to **(a)** position 1 and **(b)** position 2. The corresponding values for measurements at $111D_N$ downstream of the nozzle and $0.2y_{1/2}(x)$ above and below the centerline corresponding to **(c)** position 3; and **(d)** position 4, where $D_N$ and $y_{1/2}(x)$ are the nozzle diameter and half width of the jet, respectively.

The effectiveness of our physics-based correction is characterized by evaluating the relative error between the original and corrected laser diffraction (LD) size distributions compared to the digital inline holography (DIH) based size distribution (Figure 4). Although the average error varies from a maximum of ~70% at position 1 to ~10 % position 3, the values show significant variations with droplet size. For instance, at position 1 (Figure 4a) we see the curve monotonically decreasing from ~270% at the smallest size to nearly zero above ~250 µm. The trend in relative error continues to hold at position 2 as well, but with a reduction in the peak value to ~160% (Figure 4b). The absence of any observable LD signal in the first several bins in the off-axis positions as described earlier lead to the zero relative errors measured (Figure 4c & d).

## 4. Summary & Conclusions

In this study, we evaluate the performance of laser diffraction (LD) particle sizing of a spray from a TP6515 flat fan nozzle using a high-resolution image-based in-situ digital inline holography (DIH) measurement at four sampling regions. Apart from the direct calibration of DIH using a microruler, obviating the need for ex-situ calibration, the sampling rate used for DIH also ensures droplets are imaged only once in each image. The sampling regions on the flat fan spray were specifically chosen to highlight limitations inherent to LD due to presence of spatial variations within the spray, non-spherical droplets, and size dependent droplet velocity. The measured distributions using both techniques are clearly monomodal and highly polydisperse. However, the LD measurements and inversion led to a clear overestimation of droplet relative size distribution in the sub millimeter size range as well as an underestimation above it. This observation is quantified by variations in the geometric mean diameters of LD and DIH-inferred distributions, which are obtained by lognormal fits to results. Spatial variations in size distributions indicate that as droplet migrate downstream in a spray, the decrease in size due to breakup and evaporation, lead to a decrease in the geometric mean diameters for distributions as reported in [33] for both LD and DIH. The effect of finite liquid sheet width, as discussed in [24], leads to the decrease in the relative concentration of smaller droplets at both off-center positions and a proportionally higher reduction in droplet concentration compared to the centerline measurements.

Apart from the volumetric size distribution, DIH also enables quantification of the asphericity in the sample using the size-eccentricity joint PDF. The contours of the PDF at all four positions indicate a strong semilogarithmic scaling between diameter and eccentricity, similar to that observed in earlier measurements [33]. The observed scaling is due to the presence of droplet oscillations and rotations shown through high speed shadowgraph snapshots. Apart from a peak shift in the PDF towards smaller eccentricities with downstream distance, we also capture the broadening of the joint PDF along off-center locations. The primary cause of such a broadening is the presence of three dimensional oscillations due to oblique angles of trajectories to the wind which result in smaller relaxation time for droplets [45].

Differences in the droplet velocity with size leads to overcounting of slow-moving smaller droplets relative to larger droplets, which in part explains the observed shifts in the LD distributions. We proposed a trajectory-based correction which helps rescale the size distributions using the relative size-dependent transit time for droplets crossing the laser beam. The correction is first validated by successfully eliminating the sampling bias observed when NIST standard polydisperse beads of 50–350 μm are dropped through the LD sampling window which introduces a size-dependent sampling bias caused by the differences in settling velocities. The uncorrected measured cumulative distribution shows differences of ~30 μm at 0.50 point relative to the calibration standard. Upon application of the correction algorithm, this difference is completely removed. When the correction is extended to the spray droplets, we combine the drag model proposed for non-spherical inertial particles by [47] with the correction for liquid droplet oscillations [45]. Such a combination is new to our knowledge and has not been applied previously to correct LD measurements. We assumed droplets to be planar ellipses with the aspect ratio obtained from the ridge lines of the size-eccentricity joint PDFs. While the correction does not fully remove discrepancies between LD and DIH measurements, agreement is certainly improved especially along the centerline positions, with ~270% change in values at the smallest diameters in position 1 between the two LD measurements compared to DIH. We believe such large relative errors especially at the smaller sizes need to be carefully considered, especially for laser diffraction measurements in agricultural spray diagnostics aiming to study drift of small diameter droplets. The applied correction reduces the differences between the LD and DIH distributions over approximately two orders of magnitude but with limited effects at sizes beyond ~1 mm. Using the correction algorithm trajectory calculations, we also estimate the minimum wind speed up to which such size dependent droplet velocity will exist and beyond which the correction factor will be unity at sizes between ~40 μm and ~3500 μm at the two centerline positions examined. The estimated wind speed is ~15 m/s, a value that matches prior experimental reports by [21]

One of the reasons for the failure of the correction at larger diameters might be the limited resolution of LD; detection of larger droplets requires extremely small angle detectors, and the resolution in angle leads to an upper limit of the dynamic range. In addition, the presence of complex three dimensional oscillations (Figure 4c) cannot be fully characterized by our DIH measurement, due to our inability to make accurate measurements in the direction perpendicular to the image plane. However, such limitations can be overcome through the use of a secondary DIH imaging system placed on an orthogonal plane possibly from the top or bottom of the test section to accurately capture such complex droplet deformations. With such measurements, more accurate models for the droplet motion can be developed which may result in non-monotonic corrections.

Finally, we hypothesize that the error observed in the LD measurements originates from the previously mentioned bias toward slower moving particles. A high sample rate for laser diffraction, typically a fraction of a second [49], limits measurement independence by counting the slower moving small droplets at a greater rate than the faster moving larger droplets. The physics-based correction presented herein provides a method for correcting this bias for measurements of samples containing diameter resolved velocity gradients.

## 5. Acknowledgements

This work was supported by Winfield United. The authors acknowledge the Minnesota Supercomputing Institute (MSI) at the University of Minnesota for providing resources that contributed to the research results reported within this paper. URL: http://www.msi.umn.edu. The authors also thank Mr. Ian Marabella and Mr. Chase Christen for assistance with laser diffraction measurements and wind tunnel system operation.

**Supporting Information**

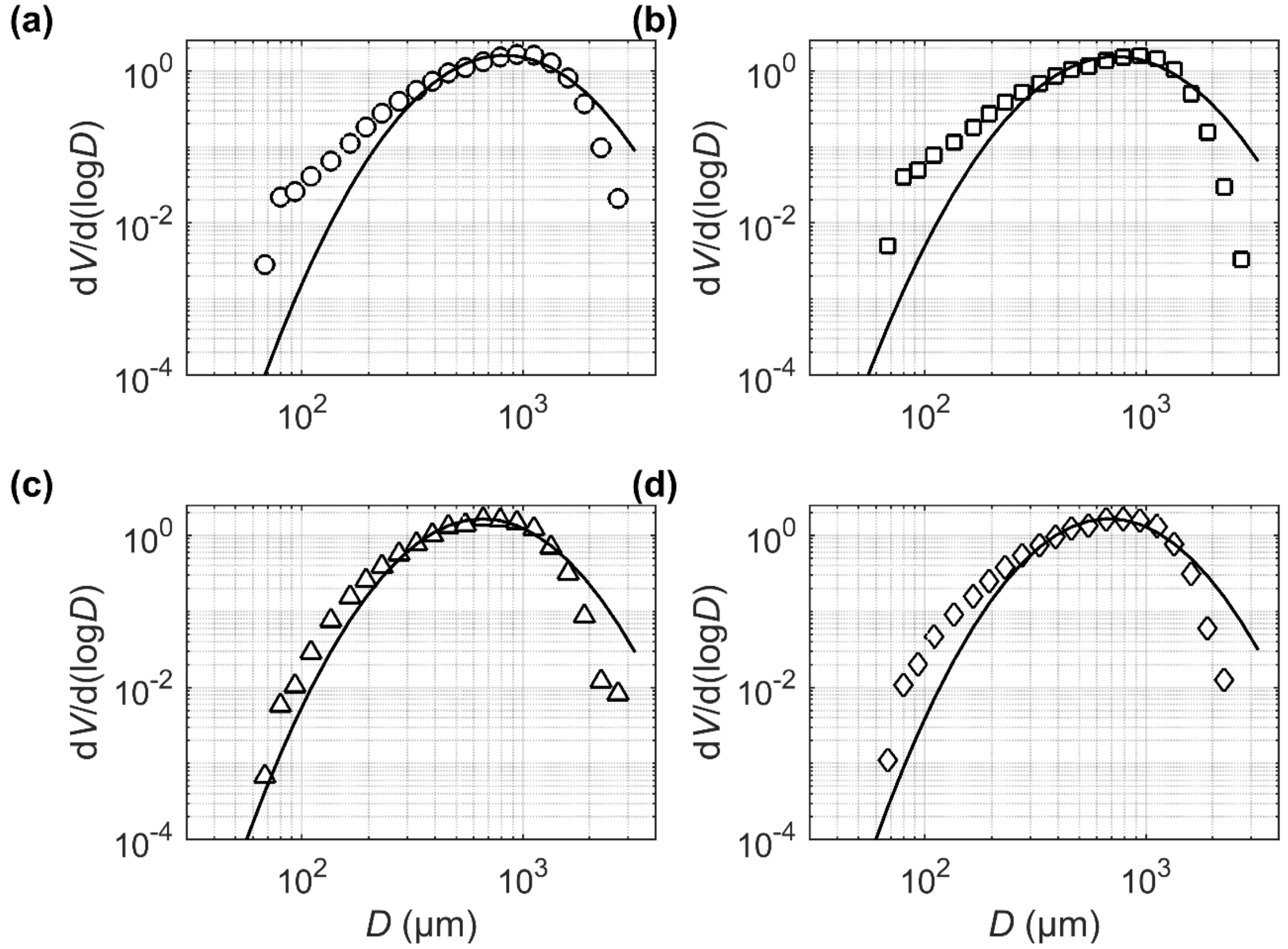

**Supplementary Figure 1** Lognormal fit to the digital inline holography (DIH) based size distribution for TP6515 flat fan nozzle spray droplets at distances of $74D_N$ and $111D_N$ downstream of the nozzle corresponding to **(a)** position 1 and **(b)** position 2. The corresponding values for measurements at $111D_N$ downstream of the nozzle and $0.2y_{1/2}(x)$ above and below the centerline corresponding to **(c)** position 3; and (**d)** position 4, where $D_N$ and $y_{1/2}(x)$ are the nozzle diameter and half width of the jet, respectively.

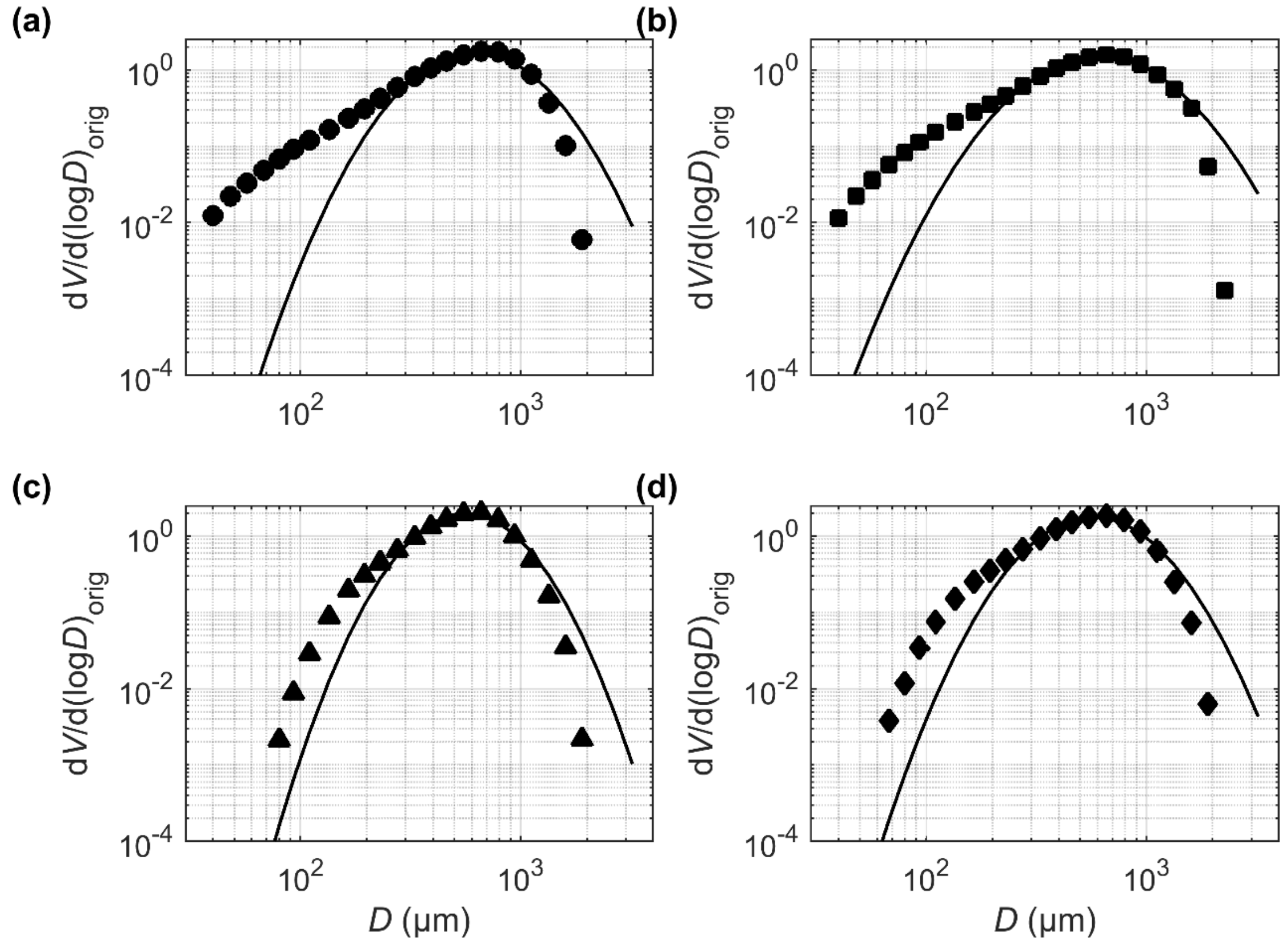

**Supplementary Figure 2** Lognormal fit to the corrected laser diffraction (LD) based size distribution for TP6515 flat fan nozzle spray droplets at distances of $74D_N$ and $111D_N$ downstream of the nozzle corresponding to **(a)** position 1 and **(b)** position 2. The corresponding values for measurements at $111D_N$ downstream of the nozzle and $0.2y_{1/2}(x)$ above and below the centerline corresponding to **(c)** position 3; and (**d**) position 4, where $D_N$ and $y_{1/2}(x)$ are the nozzle diameter and half width of the jet, respectively.

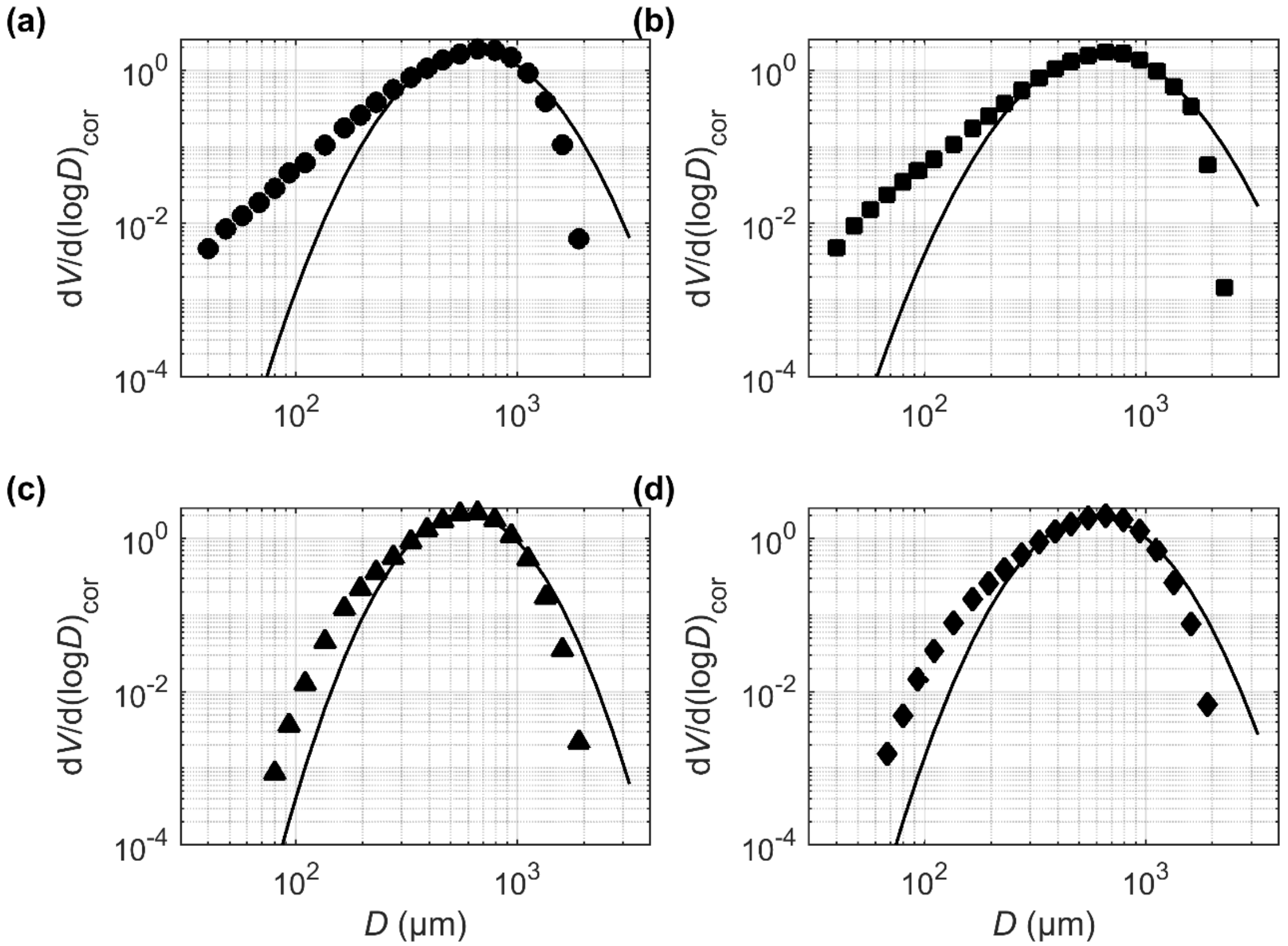

**Supplementary Figure 3** Lognormal fit to the corrected laser diffraction (LD) based size distribution for TP6515 flat fan nozzle spray droplets at distances of $74D_N$ and $111D_N$ downstream of the nozzle corresponding to **(a)** position 1 and **(b)** position 2. The corresponding values for measurements at $111D_N$ downstream of the nozzle and $0.2y_{1/2}(x)$ above and below the centerline corresponding to **(c)** position 3; and (**d)** position 4, where $D_N$ and $y_{1/2}(x)$ are the nozzle diameter and half width of the jet, respectively.

**High speed shadowgraphy:**

To visualize the complex oscillatory motion of droplets in the spray, requires a sufficiently large field of view (FOV), typically two orders larger than droplet diameter, to allow us to capture multiple cycles of oscillation from the fast moving droplet. Such a large FOV is challenging to achieve using DIH for our current experiment due to limitations in space (for optics) and laser power, but is possible with high speed shadowgraphy instead. A high intensity arc lamp is positioned on the rear window of the test section (on the side with the helium neon laser in the DIH system) and the same high speed camera and lens combination is used to capture a field of view of 76.8 mm (~$18D_N$) over 512x512 pixels (at 150 μm/pixel), recording at 25000 frames/s. Note that this field of view is still smaller than the displacement in the spanwise direction (~$70D_N$). The sampling windows are centered at the same location as position 2, 3 and 4 i.e., $111D_N$ downstream along the centerline, $0.5y_{1/2}$(x) above and below the centerline in the spanwise directions, where $D_N$ is the nozzle diameter and $y_{1/2}$(x) is the width of the spray fan, respectively (Supp. Fig. 4).

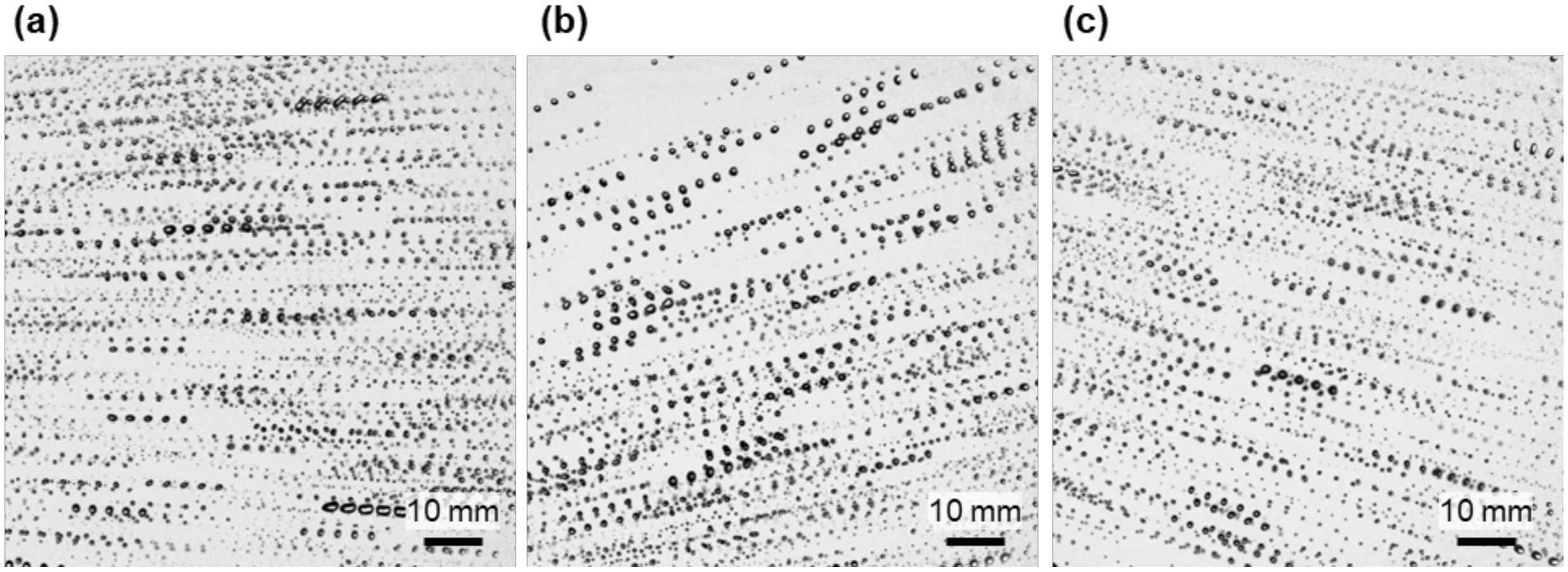

**Supplementary Figure 4** Streak images of high speed shadowgraphs after background subtraction recorded at 111$D_N$ downstream, (a) along the centerline (position 2), (b) 0.5$y_{1/2}$(x) above (position 3) and (c) below (position 4) the centerline where $D_N$ is the nozzle diameter and $y_{1/2}$(x) is the width of the spray fan, respectively. Time step between each exposure is 0.28 ms.

Figure 4 presents the temporal evolution of three specific droplets in a Lagrangian frame of reference centered on each droplet, obtained by manually tracking them over their motion across the larger field of view of the shadowgraph. As the objective of shadowgraphy was to visualize the complex oscillatory motions of predominantly larger droplets, the reduction in resolution is still capable of fully resolving these droplets.

**Supplementary video 1** High speed shadowgraphy illustrating a droplet undergoing oscillatory motion. The time separation between frames is 40 µs with scale bar of 1 mm.

**Supplementary video 2** High speed shadowgraphy illustrating a droplet undergoing rotational motion. The time separation between frames is 40 µs with scale bar of 1 mm.

**Supplementary video 1** High speed shadowgraphy illustrating a droplet undergoing three dimensional oscillatory motion. The time separation between frames is 40 µs with scale bar of 1 mm.

**Drag model for liquid droplets:**

$$C_{d,solid} = \frac{24k_s}{Re}\left(1 + 0.125(Re\, k_n/k_s)^{2/3}\right) + \frac{0.46k_n}{(1 + 5330/(Re\, k_n/k_s))}$$
$$k_n = 10^{-\alpha_2 \log(F_N)\beta_2}$$
$$k_s = (F_S^{(1/3)} + F_S^{-(1/3)})/2$$
$$\alpha_2 = 0.45 + \frac{10}{(\exp\{2.5 \log(\rho_{water}/\rho_{air})\} + 30)}$$
$$\beta_2 = 1 + \frac{37}{(\exp\{3 \log(\rho_{water}/\rho_{air})\} + 100)}$$
$$F_N = f^2 e$$
$$F_S = f e^{1.3}$$
$$f = \sqrt{1 - ecc(d_p)^2};\ e = f$$
$$C_{d,liquid} = C_{d,solid}(1 + 2.63D_c)$$
$$D_c = 2\left(1 - (1 - e(d_p))^{-0.25}\right)$$

Where $d_p$ is the droplet diameter and $e(d_p)$ is the eccentricity of the droplet from the size-shape joint PDF. The equations describing $C_{d,solid}$ is from Bagheri & Bonnadonna [1], and applies for solid non-spherical inertial particles. The correction factor for the liquid droplet $C_{d,liquid}$ is given by Mashayek & Ashgriz [2]. We combine both equations to obtain a model that describes the drag behavior for a non-spherical inertial liquid droplet as it moves through air.